\documentclass[12pt,preprint]{aastex}

\begin{document}

\title{\bf The Host Galaxies of Radio-Loud AGN:\\
The Black Hole--Galaxy Connection}

\author{Matthew O'Dowd\altaffilmark{1}}
\affil{Space Telescope Science Institute, 3700 San Martin Dr., Baltimore, MD 21218, USA; {\em odowd@stsci.edu}}
\altaffiltext{1}{also at School of Physics, University of Melbourne, 
Parkville, Victoria 3052, Australia}
\author{C. Megan Urry}
\affil{Department of Physics and Yale Center for Astronomy
and Astrophysics, P.O. Box 208121, New Haven, CT 06520-8121, USA; {\em meg.urry@yale.edu}}
\author{Riccardo Scarpa}
\affil{European Southern Observatory, Alonso de Cordova 3107, Vitacura, Casilla 19001, Santiago, Chile; {\em rscarpa@eso.org}}

\begin{abstract}
We have studied the host galaxies of a sample of radio-loud AGN
spanning more than four decades in the energy output of the nucleus.
The core sample includes 40 low-power sources (BL Lac objects) and 22 
high-power sources (radio-loud quasars) spanning the redshift range
$0.15\lesssim z\lesssim 0.5$, all imaged with the high spatial
resolution of HST. All of the sources are found to lie in luminous
elliptical galaxies, which follow the Kormendy relation for normal
ellipticals. A very shallow trend is detected between nuclear
brightness (corrected for beaming) and host galaxy luminosity.
Black hole masses are estimated for the entire sample, using both 
the bulge luminosity--black hole mass and the velocity
dispersion--black hole mass relations for local galaxies.
The latter involves a new method, using the host galaxy morphological
parameters, $\mu_e$ and $r_e$, to infer the velocity dispersion, $\sigma$,
via the fundamental plane correlation.
Both methods indicate that the entire sample of radio-loud AGN are 
powered by very massive central black holes, with 
$M_{\bullet}\sim 10^8$ to $10^{10} M_\odot$. Eddington ratios
range from $L/L_{Edd} \sim 2\times10^{-4}$ to $\sim 1$, 
with the high-power sources having higher Eddington
ratios than the low-power sources. Overall, radio-loud AGN appear to
span a very large range in accretion efficiency, 
which is all but independent of the mass of the host galaxy.
\end{abstract}


\section{Introduction}

Whether there is a link between the intrinsic power of Active Galactic Nuclei
(AGN) and their host galaxies is not known. It seems plausible that
more massive host galaxies might form in high density regions that also
support the formation of more massive nuclear black holes
\citep{Small,Haehnelt,Kauffmann} and/or that
more massive host galaxies could support an increased rate of fuelling.
For nearby, non-active galaxies there is observational evidence that the 
mass of the central supermassive black hole is correlated with bulge mass 
\citep{Magorrian,vanderMarel} and with bulge 
velocity dispersion \citep{Ferrarese,Gebhardt}.
This could lead to an observable link between emission from the region
around the black hole and the luminosity of the hosting galaxy, for
example as observed by \citet{vanderMarel99} for local spheroids.

Several studies have indeed suggested nuclear luminosity might
be related to host galaxy mass in AGN 
\citep{McLeod,Schade,Hooper}; 
however, other studies find no such relation 
\citep{Urry,McLure,Bahcall,Smith2}. 
Certainly among radio-loud AGN alone, for which the host galaxies 
are generally luminous ellipticals with well-defined morphologies,
no trend has been detected in
previous studies. This may be due to the fact that only a  
small range of (high) nuclear power has been probed in the past.

To make a clean comparison among AGN that differ only in nuclear
output (rather than host galaxy morphology, dust content, star 
formation history, etc.), we restrict the present study to radio-loud AGN
($F_{\rm 5~GHz}/F_B > 10$; Kellerman {\it et al.} 1989).
These are known to have relativistic jets formed near the central 
supermassive black hole \citep{UrryP}, and so should be governed by 
similar physics near the black hole. 
Additionally, the early-type spectral energy distributions typical of
the host galaxies \citep{Ridgway, Scarpa2, McLure2, Pentericci} make the 
K corrections straightforward.

Our goal is to probe the connection between AGN power 
(processes near the black hole) and environment (host galaxy properties) 
for radio-loud AGN over the full range of intrinsic nuclear power.
Using BL Lac objects and Radio-Loud Quasars (RLQs) to represent the
extremes of this range, it is possible to define redshift-matched
samples that span more than four orders of magnitude in intrinsic
nuclear power. This range reflects a continuum of accretion powers, 
which in turn arise from some variation in the process 
of fuelling and/or jet formation near the black hole.

\section{Matched AGN Samples and Data Corrections}

The difficulty in comparing host galaxies over a wide 
range of nuclear power lies in the redshift selection bias. 
Most low-power AGN, namely FR~I radio galaxies and Seyferts, 
are not found in complete samples beyond about $z \sim 0.2$. 
In contrast, samples with large numbers of quasars extend 
to $z \sim 0.5$ and beyond, and due to the steep luminosity
functions of quasars, few appear at low redshift.
Indeed, in any flux-limited AGN sample, there is an induced
correlation between redshift and point-source luminosity.
This can introduce apparent correlations between host galaxy 
and nuclear luminosities because galaxies evolve over even 
modest redshift ranges \citep{Bressan}.
If the redshift range is therefore restricted, the resulting
sample will ordinarily span a narrow range in nuclear power,
making it difficult to measure intrinsic trends that depend
on nuclear power.

We avoid this fundamental difficulty by selecting a low-power
sample for which intrinsic luminosity of the point source
is not well correlated with observed (selection) luminosity,
and for which samples exist to high redshift, namely, BL Lac objects. 
BL Lacs are intrinsically low-power 
blazars whose close alignment with the line of sight results in strong 
relativistic beaming of the jet emission, in many cases resulting in 
magnification of $> 1000$ times, and meaning that BL Lacs can be found 
in large numbers beyond $z \sim 1$, to the same redshift range at which 
luminous quasars can be found in large numbers. 
Their observed luminosity depends more on the Doppler factor 
than on intrinsic luminosity, thus the sample studied has no
observed correlation of (intrinsic) nuclear brightness with 
host galaxy brightness. 

We now describe the BL Lac subsample and the RLQ subsample selected
for study.

\subsection{The Low-Power Sample}

We carried out an extensive HST
imaging survey of 110 BL Lac objects with WFPC2, 
primarily in the F702W (R-band) filter.\footnote{Based 
on observations with the NASA/ESA Hubble Space Telescope, 
obtained at the Space Telescope Science Institute, which is
operated by the Association of Universities for Research in
Astronomy, Inc. under NASA contract No. NAS5-26555.}
These were a randomly selected subset of 132 BL Lacs
from six complete samples (4 X-ray-, 1 radio-, 1 optically-selected) 
spanning the full range of observed BL Lac spectral properties. 
The complete HST-observed subsample of 110 BL Lacs covers
a redshift range of $0.027 \le z \le 1.34$. 

The host galaxy parameters were extracted by fitting a model galaxy
profile plus a central point spread function (PSF) to the
azimuthally-averaged image profile. Extensive testing on simulated
data, and comparison to results from the two-dimensional analysis of a
subsample has shown that this approach allows accurate measurement of
the magnitude of both the host galaxy and nucleus and of the host
galaxy scale radius, as well as allowing us to distinguish between a
bulge- or disk-dominated galaxy profile. The excellent HST resolution
proved to be vital in determination of the morphological parameters,
as most of the critical information was within 0.5 -- 1 arcseconds of the
core. The full details of the image reduction and host galaxy fitting
can be found in \citet{Scarpa} and the host galaxy results are
presented in \citet{Urry}. 

Host galaxies were resolved in $65\%$ of the sample, with $95\%$ resolved 
for $z < 0.5$, and none resolved for $z>0.7$. 
All resolved host galaxies with sufficient signal-to-noise ratios were
well-fitted by a de Vaucouleurs profile (i.e., a bulge-dominated host),
in preference to an exponential profile (i.e., a disk-dominated
host). This strongly supports the idea that radio-loud AGN 
reside in elliptical galaxies rather than spirals.
The average K-corrected absolute magnitude of the host galaxies from
the entire HST-imaged sample is 
$M_R=-23.7 \pm 0.6$~mag (RMS dispersion).

To minimize the number of unresolved host galaxies (while still
maintaining a useful redshift range), we restrict this sample 
to $z\lesssim 0.5$ for this comparison study. We further restrict the
sample to $z\gtrsim 0.15$ 
to match the available quasar subsample (see below). The final
low-power subsample consists of 40 objects with $0.15\lesssim
z\lesssim 0.5$.

\subsection{The High-Power Sample}

For the high-power comparison sample we take RLQs 
with published imaging data comparable in quality to the BL Lac sample.
We exclude high-power radio galaxies (i.e., FR~IIs) because their 
nuclear luminosities cannot be measured directly due to 
obscuration \citep{Barthel}. Specifically, we limit the comparison
sample to quasars that satisfy the following selection criteria:

\begin{itemize}

\item
Published results from HST imaging data are available.
Our extensive testing has shown that ground-based studies 
at intermediate redshifts can result in inconsistent measurements 
of host galaxy properties. 
The stability of the HST point spread
function results in much more uniform data, and its superior 
resolution is critical for determining host morphological parameters.

\item
Restricted redshift range, $z \lesssim 0.5$.
For this redshift range our detection rate of BL Lac host galaxies 
was $95\%$, so there is minimal bias against faint hosts with 
bright nuclei. 

\item 
Host galaxies detected for full sample studied.
That is, $\sim100$\% of the RLQs with $z\lesssim 0.5$ in the published 
sample must have resolved host galaxies. As above, this avoids bias
against low host-nuclear luminosity ratios.

\end{itemize}

We identified four quasar studies meeting these criteria:
Dunlop et al. (2002), 10 objects; 
Boyce et al. (1998), 3 objects; 
Hooper et al. (1997), 6 objects; and 
Bahcall et al. (1997), 6 objects (see Table~\ref{tbl1}).
The lowest RLQ redshift is $z=0.15$, so we restrict the 
comparison sample to $0.15 \lesssim z \lesssim 0.5$. 
The final high-power sample consists of 22 RLQs
(three RLQs have duplicate observations).

The BL Lac and RLQ subsamples have indistinguishable redshift
distributions, with a Kolmogorov-Smirnov (K-S) test indicating a
$37\%$ probability of the same parent population. 
Thus the fundamental 
distinction between the two samples is nuclear luminosity.

\subsection{Photometric and Cosmological Corrections} \label{phot}

In order to compare the host galaxy properties measured 
in different studies, we first convert all results 
to the Cousin's R band,
which is closest to the HST F702W filter.
For consistency, we also used same cosmology as 
\citet{Scarpa}: $H_0 = 50$~km/s/Mpc and $q_0 = 0$. 

The spectral corrections were made using a spectral energy 
distribution (SED) derived from galaxy synthesis models 
(Bruzual \& Charlot 1993, and private communication). 
Redshifted SEDs were convolved with the transmission curves for 
both the Cousin's R and the observed bands, and then normalised to 
the published host galaxy magnitudes.
An early-type spectrum was assumed for all hosts, 
with a passively evolving population of age 8~Gyrs.
Uncertainties in the spectral corrections are 
small because we are converting between similar filters.
The uncertainties are highest in the case of the Bahcall {\it et al.}
sample, of order a tenth of a magnitude, where we are correcting
between the F606w filter (V band) and R band.
Thus variations among the particular SEDs of different 
host galaxies should not affect our results significantly.
The corrected Cousin's R absolute magnitudes for the high-power subsample
are given in Table~1. We used the average values for the two RLQs with
duplicate observations.

\section{Results}

\subsection{The Host Galaxies} \label{host}

\begin{figure}
\plotone{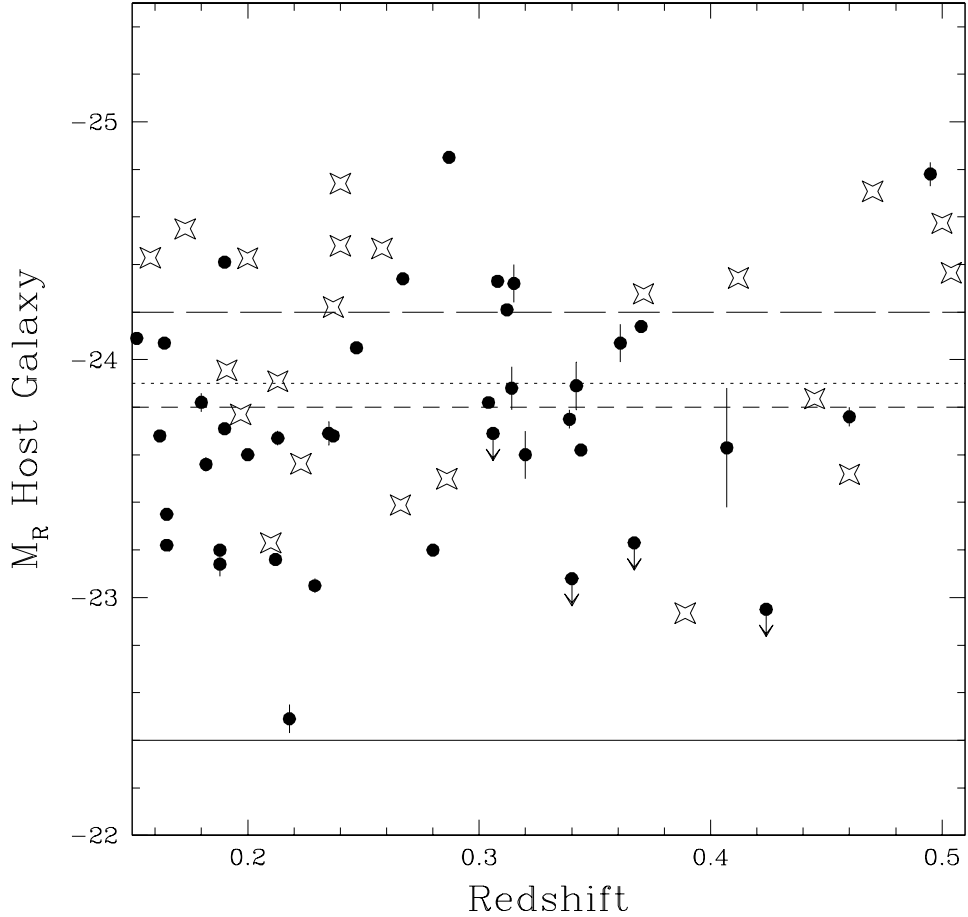}
\caption{
The K-corrected absolute R-band magnitudes of the host galaxies
of low-power radio-loud AGN ({\em circles}) and high-power
radio-loud AGN ({\em stars}) largely overlap over the redshift range
studied -- $0.15 \le z \le 0.5$. Overall, the host galaxies of 
high-power AGN are slightly brighter, with median brightness 
$M_R = -24.2 \pm 0.5$~mag ({\em long-dashed line}) compared to 
$M_R = -23.75 \pm 0.5$~mag for the low-power AGN ({\em 
short-dashed line}). A K--S test indicates a marginal difference
between the host galaxy magnitudes 
(4\% probability of being drawn from the same distribution).
The {\em solid line} shows the characteristic value for normal ellipticals
($L^* = -22.4$~mag; Efstathiou, Ellis \& Peterson 1988), while the 
{\em dotted line} shows the average magnitude of brightest cluster galaxies
($-23.9$~mag; Thuan \& Puschell 1989).
\label{fig1}
}
\end{figure}

The median K-corrected absolute magnitudes of the host 
galaxies of the subsamples differ slightly: 
$M_R = -23.75$~mag for the low-power subsample and 
$M_R = -24.2$~mag for the high-power subsample, with about half a
magnitude of scatter about their means, which are $-23.76$~mag and
$-24.02$~mag respectively.
Figure~\ref{fig1} shows host galaxy absolute magnitude plotted 
as a function of redshift. The distributions of two subsamples 
overlap completely, though the low-power subsample tends to have 
slightly fainter (but still quite luminous) host galaxies. 
A K-S test indicates that the host 
galaxy magnitudes of these two subsamples are (marginally) 
unlikely to be drawn from 
the same parent luminosity distribution (4\% probability).

In general, the host galaxies of the radio-loud AGN in this sample
have similar luminosities to brightest cluster galaxies
($M_R = -23.9$~mag; Thuan \& Puschell 1989). At their faintest they
are $\sim0.5$~mag above $L^*$ ($-22.4$~mag; Efstathiou, Ellis \&
Peterson 1988). This places them within normal bounds for non-active
elliptical galaxies, 
albeit at the bright end of the luminosity distribution. 
Every host galaxy in this sample for which the morphological
parameters were measured ($84\%$ of the sample) was
preferentially fit (and well fit) by a de Vaucouleur's 
profile, suggesting that they are in fact elliptical galaxies.

Normal (non-active) ellipticals exhibit a tight relationship between 
the log of effective radius ($r_e$) and surface brightness at that
radius ($\mu_e$), the so-called Kormendy relation 
($\mu_e = A~\log_{10}r_e + C$; Kormendy 1977), 
a projection of the Fundamental Plane.
The slope and intercept of this relation depend on
stellar dynamics (via the velocity dispersion, which is 
not known for this sample) and 
galaxy shape, size, and luminosity (which are measured).
If the Kormendy relations for this sample are similar to that of normal
elliptical galaxies, then these host galaxies are likely to be
dynamically similar to normal galaxies.

\begin{figure}
\plotone{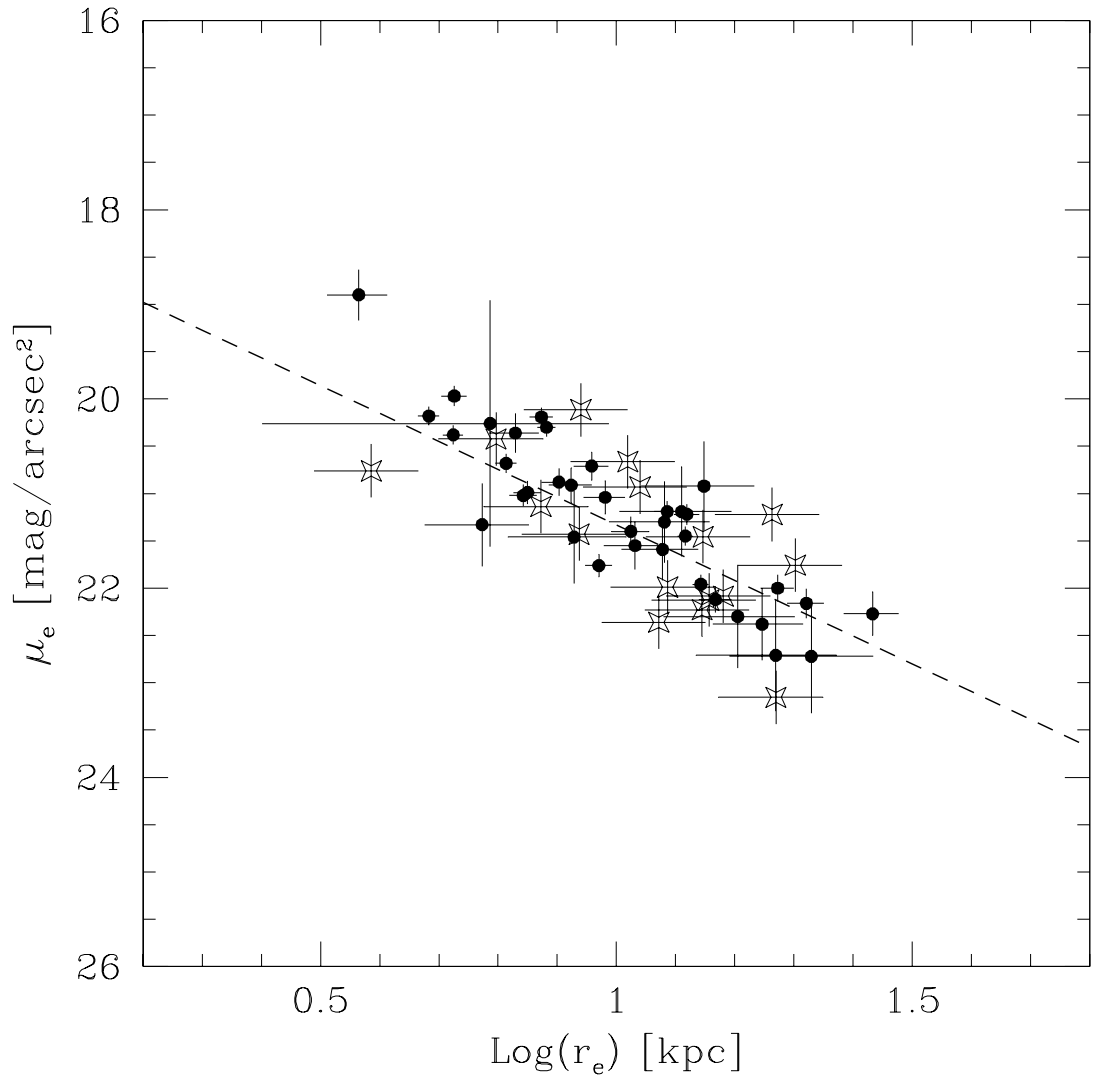}
\caption{
Surface brightness versus effective radius for the host
galaxies of low-power radio-loud AGN ({\em circles}) and high-power
radio-loud AGN ({\em stars}). The best-fit Kormendy relations for the two
samples are consistent with each other, and 
both are consistent with the Kormendy relation derived by
\citet{Hamabe} for normal elliptical galaxies ({\em dashed line}).
This suggests that the radio-loud AGN host galaxies in this sample are
dynamically similar to normal elliptical galaxies.
\label{fig2}
}
\end{figure}

Figure~\ref{fig2} shows the Kormendy relations for the host galaxies
of the two subsamples (only 16 RLQs had published effective radii).
Performing a two-dimensional K-S test \citep{Fasano}, we find that 
these subsamples are consistent with having been drawn from the
same parent distribution ($40\%$ probability). The best-fit linear
relations are $\mu_e = (3.6 \pm 0.8)~\log_{10}~(r_e / {\rm kpc}) + 
(17.4 \pm 0.7)$~mag/arcsec$^2$.
for the low-power subsample, and
$\mu_e = (2.75 \pm 1.2)~\log_{10}~(r_e / {\rm kpc}) + (18.6 \pm 1.0)$~mag/arcsec$^2$
for the high-power subsample.
The quoted errors are the one-sigma confidence levels for the 
linear fit. No published errors were available for the high-power
subsample, so a conservative $0.2~$mag/arcsec$^2$ and $20\%$ were used for $\mu_e$
and $r_e$ respectively when calculating the linear fits. These two
relations are consistent within error, and remain so even if we assume
much smaller errors in $\mu_e$ and $r_e$ for the high-power sample.

The R-band Kormendy relation for normal ellipticals in the local
universe reported by \citet{Hamabe} is $\mu_e = 2.94~\log_{10}~r_e +
18.4$, consistent within the errors with both the
low- and high-power subsamples, especially if
we take into account the evolution of the intercept with redshift
($-0.3$ to $-0.4$~mag/arcsec$^2$ at the median sample redshift of $z \sim 0.25$;
Treu {\it et al.} 2001).

\subsection{The Host Galaxy -- Nucleus Link} \label{hostnuc}

Comparison of extended radio power between BL Lac objects and 
RLQs verifies their difference in intrinsic power. 
Extended radio power provides rough bolometer of time-integrated 
power of the nucleus, independent of relativistic beaming 
of the jet emission. 
The median extended radio power of the low-power sources is 
$\log_{10}~P_{5GHz} = 24.3$~W~Hz$^{-1}$, 
compared to $\log_{10}~P_{\rm 5GHz} = 26.7$~W~Hz$^{-1}$
for the high-power sources: a difference of 2.4 orders of magnitude. 
Over four orders of magnitude separate the least radio-powerful BL Lacs 
from the most radio-powerful RLQs.
Yet host galaxy luminosities span less than one order of magnitude
($\sim 2$~mag).

Extended radio power is partly dependent on the age of the source
and on the properties of the host galaxy through which the 
radio-emitting jet must pass.
A more direct measure of power of the nucleus is the total luminosity 
emitted by the nucleus, which can sometimes be estimated from 
its observed optical brightness.
However, in BL Lac objects, strong relativistic beaming 
(which allowed us to select this matched redshift sample) 
enhances the perceived brightness by $\sim\delta^3$,
where $\delta$ is the kinematic Doppler factor of the emitting
jet plasma. To correct for beaming we use published estimates of 
Doppler factors from two blazar studies:
\citet{Ghisellini}, in which lower 
limits on the Doppler factor are calculated from measurements of bulk 
motion in the radio jet, and \citet{Dondi}, 
in which lower limits to $\delta$ are calculated 
from measurements of the ratio of $\gamma$- to X-ray photons, assuming 
a synchrotron self-Compton model.

These two studies give lower limits of Doppler factors 
for eight of our 40 BL Lac objects.
To estimate the nuclear brightnesses of the remaining sources we use
the median value of $\delta = 3.7$. This is both the median of 
the eight BL Lacs in this sample and also the median of the measured Doppler
factors for the original HST-imaged sample of 110 BL Lacs,
so is likely to be representative of the class.
Although beaming may affect the luminosities of the RLQ
sample to some degree, its effect will be 
small in comparison to the BL Lac objects
given the dominance of the thermal emission associated with
the accretion disk in RLQs.

\begin{figure}
\plotone{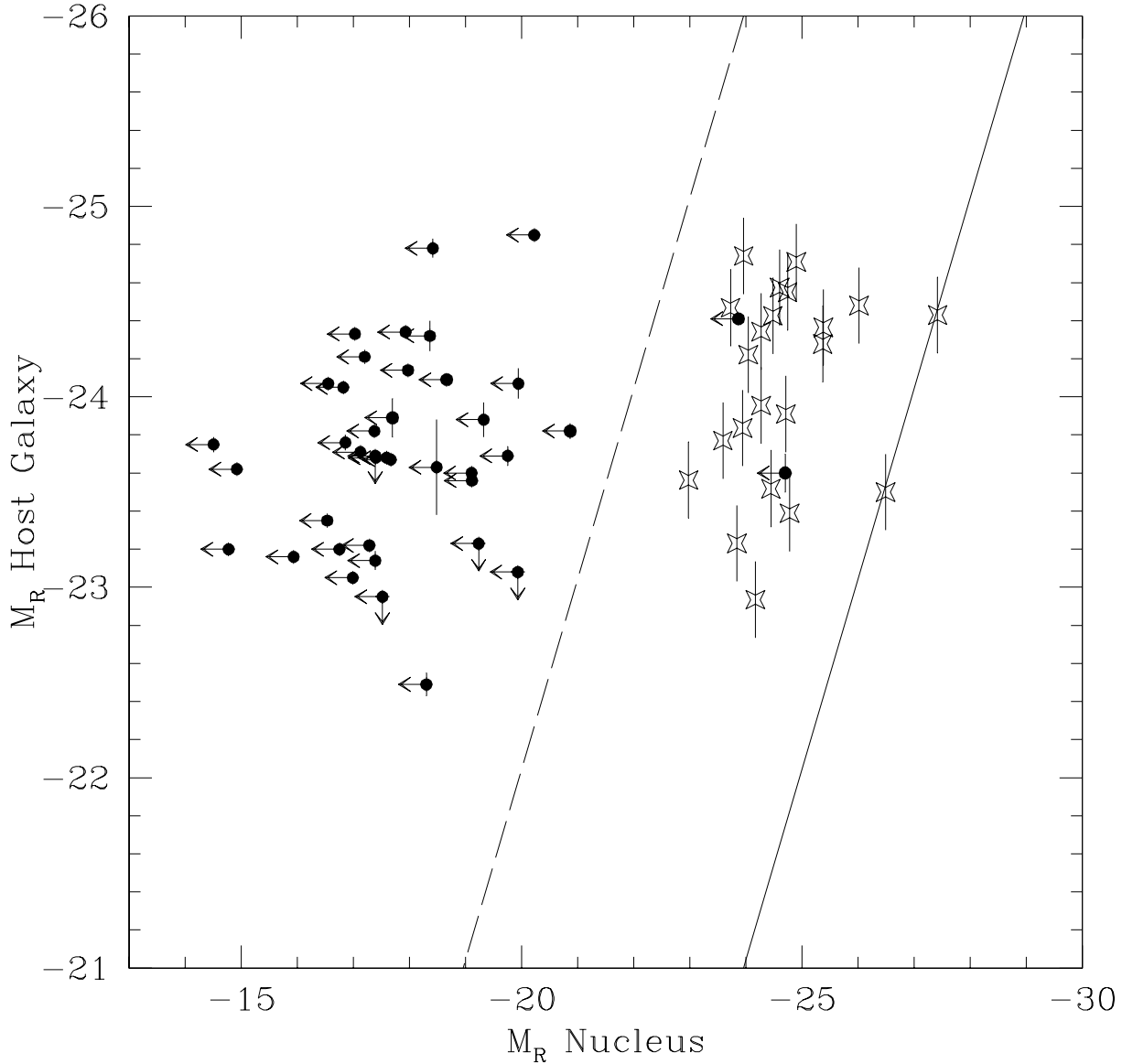}
\caption{
Absolute R-band host galaxy versus nuclear magnitude (K-corrected and,
for BL Lac nuclei, also corrected for beaming) for low-power
radio-loud AGN ({\em circles}) and high-power radio-loud AGN ({\em stars}). 
The scatter in the host galaxy 
magnitudes is small (RMS is 0.6~mag), compared to more than 4 orders 
of magnitude in nuclear luminosity between the least luminous BL Lac 
nuclei and the most luminous RLQs. There is a statistically
significant trend between host galaxy and nuclear magnitude,
but it is considerably shallower than the line of constant
Eddington ratio ({\em solid line} is $L/L_{Edd}=1$,
{\em dashed line} is $L/L_{Edd}=0.01$) obtained by assuming the 
host galaxy luminosity--black hole mass relation reported by 
\cite{KormendyG}. 
\label{fig3}
}
\end{figure}

Figure~\ref{fig3} shows absolute host galaxy magnitude
versus absolute nuclear magnitude for the low- and high-power subsamples.
The BL Lac nuclear magnitudes have been K corrected and corrected
for beaming, with the lower limits in the Doppler factors translating 
to upper limits in nuclear luminosities. The median absolute nuclear
magnitude for the 8 BL Lacs with measured 
Doppler factor limits is $M_R \gtrsim -19.23$~mag, while for the 
entire low-power subsample it is $M_R \gtrsim -17.59$~mag. 
For the high-power sources (K corrected only) the
median absolute magnitude is $-24.5$~mag. 
At least four orders of magnitude separate the least powerful BL Lacs 
($M_R \gtrsim -17$~mag) and the most luminous RLQs ($M_R \sim -27$~mag). 

Although the host galaxies of the low-power sample span a similar
range in magnitude to those of the high-power sample, they are on
average slightly fainter. This leads to a shallow trend between the
host galaxy and nuclear luminosities across the combined sample. We
calculated the Kendall's $\tau$ correlation coefficient, censoring the
upper limits in the beaming-corrected BL Lac nuclear luminosities (and
for four of their host luminosities). The trend was significant for the
case where we include the entire low-power subsample, beaming-corrected
with the median Doppler factor (0.01\% probability of no correlation; 
Kendall's ($\tau$) for 62 points: 0.555), and for the case where we
include only 
those of the low-power subsample with measured Doppler factor limits (0.5\%
probability of no correlation; Kendall ($\tau$) for 30 points: 0.671).

Although this trend is statistically significant, it is also very
shallow. Performing a linear fit with the high-power subsample
combined with the non-beaming-corrected low-power subsample, we find
that the host galaxy luminosity increases by only $1$~mag for an
increase of $7$~mag in the luminosity of the nucleus. Any introduction
of beaming correction decreases this gradient further. Applying the
median Doppler factor corrections to the low-power subsample yields an
increase in $\sim 10$~mag in the nuclear luminosity for each magnitude
of host galaxy luminosity. 

The shallowness of this trend, and the narrow range in luminosity
exhibited by these host galaxies demonstrates that, for radio-loud AGN, the
luminosity (and hence mass) of the host galaxy has little relation to 
the energy output of the nucleus. If the correlation between 
bulge luminosity and central black hole mass observed in nearby, 
non-active galaxies \citep{Magorrian,vanderMarel}
also applies to these host galaxies, then this tight range of host
luminosities implies a small range of (very high) black hole masses,
and hence a very large range of Eddington ratios.
In the following sections we derive these Eddington ratios using two 
different methods for estimating the black hole masses.

\subsection{Black Hole Masses and Eddington Ratios from Bulge Luminosity}

Since the host galaxies of radio-loud AGN appear to be normal ellipticals
in every respect measured (see Sect.~\ref{host}), it is plausible that the AGN black hole
mass and host galaxy luminosity are correlated, as observed in nearby, 
non-active galaxies \citep{Magorrian,vanderMarel}. If so, radio-loud
AGN are powered by very massive central black holes. We use the
relation presented by \citet{KormendyG} to calculate these: 

\begin{equation}
M_{\bullet}=0.78\times 10^8 M_{\odot}\frac{L_{B,bulge}}{10^{10} L_{B,\odot}} ~.
\end{equation}

\noindent
Assuming an early-type spectrum to derive $L_{B,bulge}$ from the R-band 
magnitudes (see Sect.~\ref{phot}), we find median black hole masses of 
$1.1 \times 10^9 M_\odot$ (mean: $1.2 \times 10^9 M_{\odot}$)
for the low-power subsample, and 
$1.7 \times 10^9 M_\odot$  (mean: $2.0 \times 10^9 M_{\odot}$)
for the high-power subsample, with standard deviations $6.8 \times 10^8
M_\odot$ and $1.0 \times 10^9 M_\odot$ respectively.
The median for the entire sample is $M_{\bullet}=1.2 \times 10^9
M_{\odot}$, with standard deviation $7.7
\times 10^8 M_{\odot}$ (mean: $1.2 \times 10^9 M_{\odot}$).

\begin{figure}
\plotone{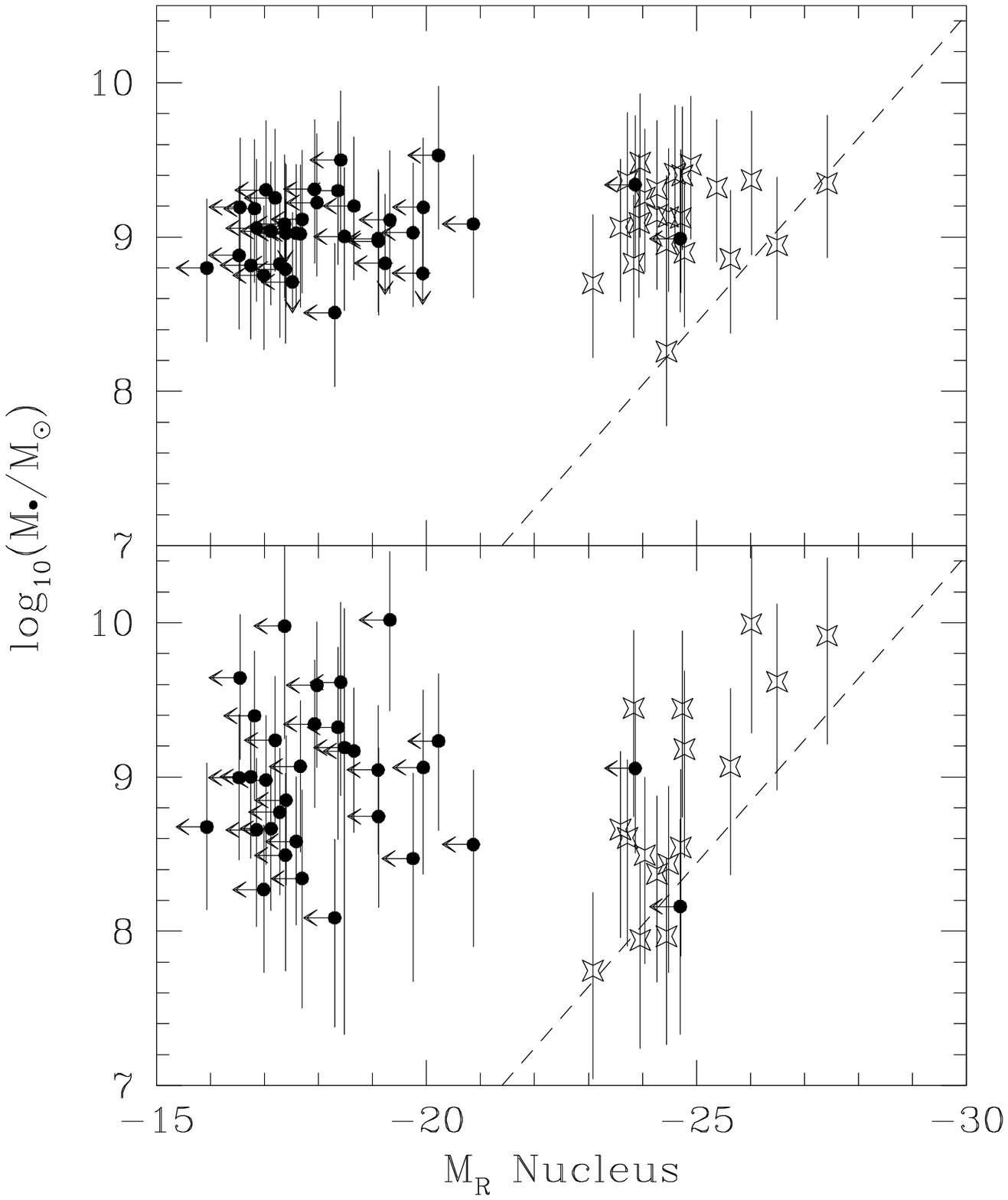}
\caption{
Derived black hole masses versus nuclear magnitude (again, K-corrected and,
for BL Lac nuclei, corrected for beaming) for low-power radio-loud AGN 
({\em circles}) 
and high-power radio-loud AGN ({\em stars}). {\em Upper plot:} using the  
bulge luminosity correlation \citep{Magorrian, KormendyG}. 
{\em Lower plot:} using the stellar velocity dispersion correlation
\citep{Gebhardt, Ferrarese, KormendyG} combined with
the Fundamental Plane relation \citep{Jorgensen}.
Both plots suggest that these radio-loud AGN exhibit a relatively
small range of
high black hole masses for a very large range ($>4$ orders of
magnitude) in energy output from the nucleus.
Although radio-loud AGN can extend to very faint nuclear magnitudes,
they appear to be cut off at the bright end by an envelope consistent 
with the Eddington luminosity ({\em dashed line}).
\label{fig4}
}
\end{figure}

The upper part of
Figure~\ref{fig4} shows the derived black hole masses versus
(corrected) magnitude of the nucleus. The error in black hole mass is
dominated by the scatter in the $M_{\bullet}$---$L_{bulge}$ relation 
(the RMS dispersion is a factor of 2.8; Kormendy \& Gebhardt 2001), 
but also affected by the uncertainty in the host galaxy magnitudes and 
in the spectral corrections (a factor of $\sim \pm 10\%$).

The shallow correlation noted between the luminosities of host galaxy
and nucleus 
(Sect.~\ref{hostnuc}) leads to a correlation between black hole mass and 
nuclear luminosity; however, when we include the error bars in the
black hole masses, the correlation is not statistically significant.

Given AGN black hole masses estimated in this way, we can derive the 
Eddington luminosity:

\begin{equation}
L_{Edd} = \frac{4\pi G M_{\bullet} m_p c}{\sigma_T} ~,
\end{equation}

\noindent
where $M_{\bullet}$ is the black hole mass, $m_p$ the mass of the proton, 
and $\sigma_T$ the Thompson cross-section. From this we calculate 
the Eddington ratio using the corrected nuclear magnitudes, assuming a  
spectral index of $\alpha=1$ to estimate bolometric luminosity.
The nuclear magnitudes of BL Lac objects are corrected for
beaming using lower limits to the Doppler factors, meaning the
calculated Eddington ratios are upper limits. 

\begin{figure}
\plotone{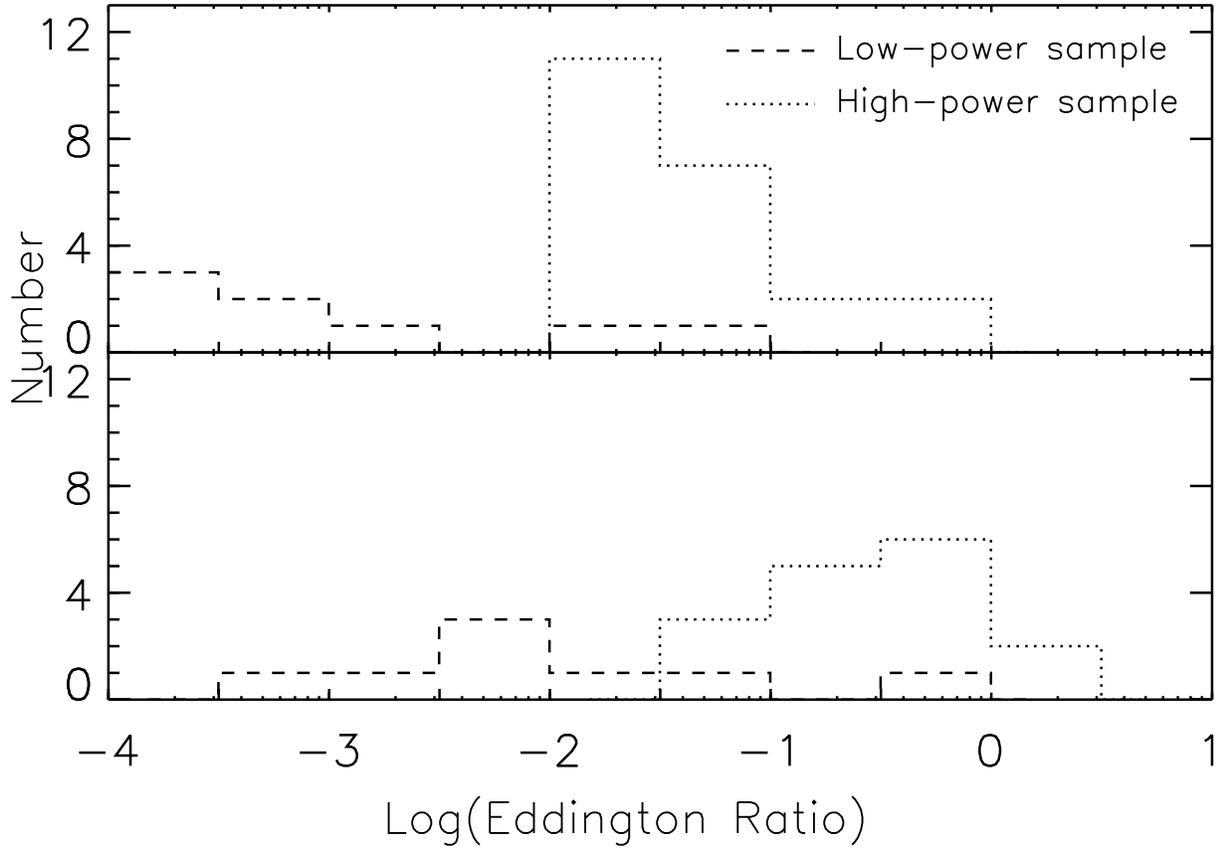}
\caption{
Histogram of Eddington ratios for low- and high-power
radio-loud AGN (including only the eight low-power sources with measured
Doppler factors). Black hole masses are calculated from the bulge luminosity
relation ({\em upper histogram}) and from the velocity dispersion relation
({\em lower histogram}).
\label{fig5}
}
\end{figure}

For the eight BL Lacs with measured Doppler factor limits, the median
Eddington ratio limit is $\frac{L_{bol.}}{L_{Edd}} \lesssim
0.002$. For the entire low-power sample, corrected with the median
Doppler factor, we obtain 
$\frac{L_{bol.}}{L_{Edd}} \lesssim 3\times10^{-4}$. The median
Eddington ratio for the high-power sources is much higher, with 
$\frac{L_{bol.}}{L_{Edd}} = 0.1$.
The upper part of Figure~\ref{fig5} shows the histogram of Eddington
ratios (or Eddington ratio limits) for the reduced low-power sample
and the high-power sample.

\subsection{Black Hole Masses and Eddington Ratios from Velocity Dispersion Estimates}

Black hole mass appears to be much more tightly correlated 
with velocity dispersion ($\sigma_e$) than with bulge luminosity 
\citep{Gebhardt, Ferrarese, KormendyG}.
If we assume that our host galaxies lie on the Fundamental Plane for
normal ellipticals, we can infer velocity dispersions using the
measured values of $r_e$ and $\langle I \rangle_e$ --- the mean 
surface brightness within $r_e$. This assumption is reasonable, as we 
know that these host galaxies are indistinguishable from normal
elliptical galaxies morphologically, and from \S~\ref{host} we
know that they are consistent dynamically, with their Kormendy
relations matching.

We use the Fundamental Plane parameters measured by \citet{Jorgensen}
from R-band photometry of a large sample of E and S0 galaxies across
several clusters:

\begin{equation}
\log r_e = 1.24 \log \sigma_e - 0.82~\langle I \rangle_e + \gamma ~.
\end{equation}

\noindent
The evolution of the zero point, $\gamma$, with redshift is calculated
using the measurements of Fundamental Plane parameters at different
redshifts by \citet{Jorgensen2}, yielding
$\gamma = 0.2132 z - 1.31 \times 10^{-3}$.

The velocity dispersions derived were used to calculate black hole
masses using the relation:

\begin{equation}
M_{\bullet}=1.3 \times 10^8 M_{\odot} (\frac{\sigma_e}{200~{\rm km~s}^{-1}})^{4.2} ~.
\end{equation}

The value of the exponent is somewhat uncertain, with conflicting 
measurements in the literature. \citet{KormendyG} measure an $\alpha=3.65$, 
while \citet{Merritt} measure $\alpha=4.72$. To reflect this uncertainty, 
we adopt a mean value of $\alpha=4.2$ with an uncertainty equal to the standard 
deviation of the two values: $\pm 0.75$.

The median black hole masses derived using this method are, for the 
low-power sample: $M_{\bullet}=1.1 \times 10^9 M_{\odot}$ 
(mean: $1.8 \times 10^9 M_{\odot}$); and for the high-power sample: 
$M_{\bullet}=4.6 \times 10^8 M_{\odot}$ (mean: $2.0 \times 10^9 M_{\odot}$).
Black hole masses derived with this method show a greater spread than
those derived using the $M_{\bullet}$---$L_{bulge}$ relation, with
standard deviations $2.3 \times 10^9 M_{\odot}$ and 
$3.0 \times 10^9 M_{\odot}$ respectively.
The median for the entire sample is $M_{\bullet}=1.0 \times 10^9
M_{\odot}$(mean: $8.9 \times 10^8 M_{\odot}$), with standard deviation $2.5
\times 10^9 M_{\odot}$ .

The lower part of Figure~\ref{fig4} shows black hole mass versus the 
magnitude of the nucleus for the masses derived in this
section.
The errors in the black hole masses are dominated by the scatter in the
Fundamental Plane relation, which is around 25\% in
$\sigma_e$ for a given $\mu_e$ and $r_e$. 
The uncertainty in $\mu_e$ and
$r_e$ affect the errors to a lesser extent, and the errors in the
actual Fundamental Plane fit parameters are small in comparison. 
The assumed error in $\alpha$, although leading to differences of 
only $\sim \pm15\%$ in black hole mass, compounds with the error in
$\sigma_e$. As a result, although the scatter in the 
$M_{\bullet}$---$\sigma_e$ relation is small, the final errors in
the black hole masses calculated by this method are of similar order
to those calculated from the $M_{\bullet}$---$L_{bulge}$ relation.
Again, taking these errors into account, there is no statistically
significant trend between black hole masses derived in this section 
and nuclear luminosity, beaming corrected or otherwise. 

The median Eddington ratio limit for the eight BL Lacs with measured Doppler
factor limits is $\frac{L_{bol.}}{L_{Edd}} \lesssim 0.01$, 
while the median for the entire low-power subsample is
$\frac{L_{bol.}}{L_{Edd}} \lesssim 3\times10^{-4}$. 
Again, these Eddington ratios are a great deal smaller than the median
for the high-power sample calculated by this method:
$\frac{L_{bol.}}{L_{Edd}} \lesssim 0.4$.
The lower part of figure~\ref{fig5} shows the histogram of the Eddington
ratios calculated in this section. 

\section{Discussion}

The results from the two methods of black hole mass calculation agree:
radio-loud AGN are powered by very massive central black holes,
typically with $M_{\bullet} > 10^8 M_{\odot}$, and more often with 
$M_{\bullet} \sim 10^9 M_{\odot}$.

These results are generally consistent with the black hole masses 
derived for luminous radio-loud AGN in other studies. \citet{Falomo} find masses in 
the range $5\times 10^7  M_{\odot}$ to  $9 \times 10^8 M_{\odot}$ for 
their sample of 7 low-redshift ($z < 0.055$) BL Lacs, calculated from
spectroscopic measurements of stellar velocity dispersion. These 
fall within the range of masses found for our sample of BL
Lacs via the velocity dispersion relation, although they tend 
toward the low-mass end of that range. This may indicate an
evolutionary or selection effect between the two epochs studied.
\citet{McLure3} find black hole masses in the range $2\times 10^8
M_{\odot}$ to $2\times 10^9 M_{\odot}$ for a sample of 22
radio-loud quasars, determined through reverberation
mapping. These are consistent with the masses found for our high-power
sample. 

Our results support the idea that, although some radio-quiet
objects may host very massive central black holes \citep{Dunlop},
luminous radio sources may require them. Dunlop et al. suggest a black
hole mass threshold for radio-loud AGN of $5 \times 10^8 M_{\odot}$,
which is supported by the black hole masses derived from bulge
luminosity in this study. Masses derived in this study via the
$M_{\bullet}$---$\sigma_e$ relation suggest a threshold of 
$\sim 1\times 10^8 M_{\odot}$. Whatever its value, we find that neither
the threshold nor the distribution of black hole masses in radio-loud AGN
depends on the actual level of radio emission, within the range 
represented by these radio sources, or on
the overall energy output of the nucleus.

As a consequence, radio-loud AGN exhibit an extremely broad range of accretion
rates; from 
$\frac{L_{bol.}}{L_{Edd}} \lesssim 2\times 10^{-4}$ to
$\frac{L_{bol.}}{L_{Edd}} \sim 1.0$.
Across this range the host galaxies 
span a remarkably tight range of high stellar luminosities --- all
within one magnitude of brightest cluster galaxies --- suggesting that,
at least for radio-loud AGN, there is at most a very weak relation between
the properties of the host galaxy and both the overall rate and the
efficiency of fuelling of the black hole.

\section{Conclusions}

We find that the host galaxies of radio-loud AGN are 
luminous ellipticals, occupying the low surface-brightness tail of the
Kormendy relation for normal elliptical galaxies, and are statistically
consistent with this relation.
Comparing the host galaxies of low-power and high-power radio-loud AGN,
we find general overlap, with a slight difference in median absolute
Cousins R magnitudes, $-23.75$~mag  and $-24.2$~mag, respectively. 
After correcting the (highly beamed) low-power AGN for Doppler beaming, 
we find a significant positive trend between nuclear 
and host galaxy luminosity, but
with a very shallow slope --- a factor of 1.3 in host galaxy
brightness over at least four orders of magnitude in nuclear
luminosity --- ruling out a close relation between host galaxy and
nuclear luminosity in radio-loud AGN.

We find that the central black holes of luminous radio-loud AGN are universally
large, with median black hole mass $\sim 10 \time 10^9 M_\odot$ for
this sample.
This is found to be the case using either the $M_{\bullet}$---$L_{bulge}$
relation and the $M_{\bullet}$---$\sigma_e$
relations to derive black hole masses. This supports the view that 
a high central black hole mass is an important factor in generating 
a powerful radio source.

No correlation is found
between black hole mass and energy output from the nucleus. 
Rather, the black hole masses derived span a surprisingly small range
compared to the range in intrinsic power of this sample. 
Eddington ratios for radio-loud AGN span more than four
orders of magnitude, with $\frac{L_{bol.}}{L_{Edd}} \lesssim 2\times
10^{-4}$ in the lowest-power sources to 
$\frac{L_{bol.}}{L_{Edd}} \sim 1$ in the highest. 
Across this range, the host galaxies luminosities are tightly
constrained, all within one magnitude of brightest cluster galaxies. 
Thus, although the
properties of the host galaxy may have a strong influence the mass of
its central black hole, they have at most a very weak influence on
the mass accretion rate in radio-loud AGN.


\acknowledgments
We thank Aldo Treves and Laura Maraschi for very helpful discussions.
Support for this work was provided by NASA through grant numbers
GO-05938.01-94A, GO-05939.01-94A, GO-06363.01-95A and GO-07893.01-96A 
from the Space Telescope Science Institute, which is operated by 
AURA, Inc., under NASA contract NAS5-26555.

\clearpage

\begin{deluxetable}{lccccc}
\tablecaption{The High-Power Sample\tablenotemark{*}\label{tbl1}}
\tablewidth{0pt}
\tablehead{
\colhead{Name}         & \colhead{z} & 
\colhead{M$_R$(host)}  & \colhead{M$_R$(nucl.)} & 
\colhead{R$_e$(kpc)}  & \colhead{P$_{\rm 5GHz}$($\log$ W Hz$^{-1}$)}}
\startdata
\multicolumn{6}{l}{~~~{\em \citet{Dunlop}}}\\
0137+012   &0.258 &-24.47 &-23.50 &15.16 &26.78\\
0736+017   &0.191 &-23.95 &-24.05 &13.96 &26.81\\
1004+130   &0.240 &-24.58 &-25.75 &8.71  &26.40\\
2141+175   &0.213 &-23.91 &-24.49 &8.65  &26.33\\
2247+140   &0.237 &-24.22 &-23.82 &14.34 &26.77\\
2349-014   &0.173 &-24.66 &-24.04 &20.06 &26.24\\
1020-103   &0.197 &-23.77 &-23.37 &7.46  &26.13\\
1217+023   &0.240 &-24.74 &-23.74 &11.80 &26.45\\
2135-147   &0.200 &-24.40 &-23.40 &12.20 &26.67\\
2355-082   &0.210 &-23.23 &-23.62 &10.97 &25.96\\
\multicolumn{6}{l}{~~~{\em \citet{Boyce}}}\\
0202-760   &0.389 &-22.93 &-23.95 & 3.85 &27.30\\
3C351      &0.371 &-24.28 &-25.15 & 6.26 &26.90\\
0312-770   &0.223 &-23.56 &-22.76 &18.61 &26.43\\
\multicolumn{6}{l}{~~~{\em \citet{Hooper}}}\\
1138+0003  &0.500 &-24.57 &-24.38 & ---  &25.48\\
1218+1734  &0.445 &-23.84 &-23.72 & ---  &25.20\\
1222+1235  &0.412 &-24.34 &-24.05 & ---  &25.06\\
1230-0015  &0.470 &-24.71 &-24.68 & ---  &25.65\\
2348+0210  &0.504 &-24.36 &-25.16 & ---  &25.45\\
2351-0036  &0.460 &-23.52 &-24.22 & ---  &26.27\\
\multicolumn{6}{l}{~~~{\em \citet{Bahcall}}}\\
1004+130   &0.240 &-24.37 &-25.84 & 9.37 &27.26\\ 
3C273      &0.158 &-24.43 &-27.20 &18.33 &28.41\\
1302-102   &0.286 &-23.50 &-26.27 &10.45 &26.40\\
3C323.1    &0.266 &-23.39 &-24.56 &14.01 &26.95\\
2135-147   &0.200 &-23.45 &-25.12 &16.63 &27.02\\
2349-014   &0.173 &-24.44 &-25.01 &26.23 &26.40\\
\enddata
\tablenotetext{*}{These values have been converted from the published 
values to our adopted cosmology, $H_0 = 50 $~km/s/Mpc and $q_0 = 0$, 
and in the case of magnitudes, to Cousin's R band.}
\end{deluxetable}

\clearpage

\clearpage

\clearpage

\clearpage

\clearpage

\clearpage


\clearpage

\begin{thebibliography}{}
\small
\bibitem[Bahcall {\it et al.}(1997)]{Bahcall}	Bahcall, J. N., Kirhakos, S., Saxe, D. H. \& Schneider. D. P. 1997 ApJ 479, 642
\bibitem[Barthel(1989)]{Barthel}		Barthel, P. 1989 SciAm 260, 20
\bibitem[Boyce {\it et al.}(1998)]{Boyce}	Boyce, P. J., Disney, M. J., Blades, J. C., {\it et al.} 1998 MNRAS 298, 121
\bibitem[Bressan, Chiosi \& Fagotto(1994)]{Bressan}	Bressan, A., Chiosi, C. \& Fagotto, F. 1994 ApJs 94, 63
\bibitem[Bruzual \& Charlot(1993)]{Bruzual}	Bruzual \& Charlot 1993 ApJ 405, 538
\bibitem[Dondi \& Ghisellini(1995)]{Dondi}	Dondi, L. \& Ghisellini, G. 1995 MNRAS 273, 583
\bibitem[Dunlop et al. (2002)]{Dunlop}		Dunlop, J. S., McLure, R. J., Kukula, M. J., et al. 2002 MNRAS, in press
\bibitem[Efstathiou, Ellis \& Peterson(1988)]{Efstathiou}	Efstathiou, G., Ellis, R. S. \& Peterson, B. A. 1988 MNRAS 232, 431
\bibitem[Fasano \& Franceschini(1987)]{Fasano}	Fasano, G.\& Franceschini, A. 1987 MNRAS 225, 155
\bibitem[Falomo, Kotilainen \& Treves(2002)]{Falomo}	 Falomo, R., Kotilainen, J. K. \& Treves, A. 2002 ApJ Letters, in press
\bibitem[Ferrarese \& Merritt(2000)]{Ferrarese}	Ferrarese, L. \& Merritt, D. 2000 ApJ 539, L9
\bibitem[Gebhardt {\it et al.}(2000)]{Gebhardt}	Gebhardt, K., Bender, R., Bower, G., {\it et al.} 2000 ApJ 539, L13
\bibitem[Ghisellini {\it et al.}(1993)]{Ghisellini}	Ghisellini, G., Padovani, P., Celotti, A. \& Maraschi, L. 1993 ApJ 407, 65
\bibitem[Haehnelt \& Rees(1993)]{Haehnelt}	Haehnelt, M. G. \& Rees, M. J. 1993 MNRAS 263, 168
\bibitem[Hamabe \& Kormendy(1987)]{Hamabe}	Hamabe, M. \& Kormendy, J. 1987 in IAU Symp. 127, Structure and Dynamics of Elliptical Galaxies, ed. P. T. de Zeeuw (Dordrecht: Kluwer), 379
\bibitem[Ho(1999)]{Ho}				Ho, L. 1999 in Observational Evidence for Black Holes in the Universe, ed. S. K. Chakrabarti (Dordrecht: Kluwer), 157
\bibitem[Hooper, Impey \& Foltz(1997)]{Hooper}	Hooper, E. J., Impey, C. D. \& Foltz, C. B. 1997 ApJ 480, L98
\bibitem[Jorgensen, Marijn \& Kjaegaard(1995)]{Jorgensen} Jorgensen, I., Marijn, F. \& Kjaegaard, P. 1995 MNRAS 276, 1341
\bibitem[Jorgensen {\it et al.}(1999)]{Jorgensen2}	Jorgensen, I., Marijn, F., Jens, H. \& van Dokkum, P. G., 1999 MNRAS 308, 833
\bibitem[Kauffmann \& Haehnelt(2000)]{Kauffmann}	Kauffmann, G. \& Haehnelt, M. G. 2000 MNRAS 311, 576
\bibitem[Kellerman {\it et al.}(1989)]{Kellerman}	Kellerman, K. I., Sramek, R., Schmidt, {\it et al.} 1989 AJ 98, 1195
\bibitem[Kormendy(1977)]{Kormendy}		Kormendy, J. 1977 ApJ 218, 333
\bibitem[Kormendy \& Gebhardt(2001)]{KormendyG}	Kormendy, J. \& Gebhardt, K. 2001 in The 20th Texas Symposium on Relativistic Astrophysics, ed H. Martel \& J. C. Wheeler (AIP)
\bibitem[Magorrian {\it et al.}(1998)]{Magorrian}	Magorrian, J., Tremaine, S., Richstone, D. {\it et al.} 1998 AJ 115, 228
\bibitem[McLeod, Rieke \& Storrie-Lombardi(1999)]{McLeod}		McLeod, K. K., Rieke, G. H. \& Storrie-Lombardi, L. J. 1999 ApJ 511, L67
\bibitem[McLure, Kukula \& Dunlop(1999)]{McLure}	McLure, R. J., Kukula, M. J. \& Dunlop, J. S. 1999 MNRAS 308, 377
\bibitem[McLure, Dunlop \& Kukula(2000)]{McLure2}	McLure, R. J., Dunlop, J. S. \& Kukula, M. J. 2000 MNRAS 318, 693
\bibitem[McLure \& Dunlop(2001)]{McLure3}	McLure, R. J. \& Dunlop, J. S.  2002 MNRAS, in press
\bibitem[Merritt \& Ferrarese(2001)]{Merritt}	Merritt, D. \& Ferrarese, L. 2001 MNRAS.320, L30
\bibitem[Padovani \& Giommi(1995)]{Padovani}	Padovani, P. \& Giommi, P. 1995 ApJ 444, 567
\bibitem[Pentericci {\it et al.}(2001)]{Pentericci}	Pentericci, L., McCarthy, P. J., R\"ottgering, H. J. A., Miley, G. K., van Breugel, W. J. M. \& Fosbury, R. 2001 ApJS 135, 63
\bibitem[Ridgway \& Stockton(1997)]{Ridgway}	Ridgway, S. E., Stockton, A. 1997 AJ 114, 511
\bibitem[Scarpa {\it et al.}(2000)]{Scarpa}		Scarpa, R., Urry, C. M., Falomo, R., Pesce, J. \& Treves, A. 2000 ApJ 532, 740
\bibitem[Scarpa {\it et al.}(2000a)]{Scarpa2}	Scarpa, R., Urry, C. M., Padovani, P., Calzetti, D., O'Dowd, M. 2000 ApJ 544, 258
\bibitem[Schade, Boyle \& Letawsky(2000)]{Schade}	Schade, D. J., Boyle, B. J. \& Letawsky, M. 2000 MNRAS, 315 \& 498
\bibitem[Small \& Blandford(1992)]{Small}	Small \& Blandford 1992 MNRAS 259, 725
\bibitem[Smith \& Heckman(1990)]{Smith}		Smith, E. P. \& Heckman T. M. 1990 ApJS 69, 365
\bibitem[Smith {\it et al.}(1986)]{Smith2}      Smith, E. P., Heckman, T. M., Bothun, G. D., Romanishin, W. \& Balick, B. 1986 ApJ 306, 64
\bibitem[Thuan \& Puschell(1989)]{Thuan}	Thuan, T. X. \& Puschell, J. J. 1989 ApJ 346, 34
\bibitem[Treu {\it et al.}(2001)]{Treu}         Treu, T., Stiavelli, M., Bertin, G., {\it et. al} 2001 MNRAS 326, 237
\bibitem[Urry \& Padovani(1995)]{UrryP}		Urry, C. M. \& Padovani, P. 1995, PASP 107, 803
\bibitem[Urry {\it et al.}(2000)]{Urry}		Urry, C. M., Scarpa, R., O'Dowd, M., {\it et al.} 2000, ApJ 532, 816
\bibitem[van der Marel(1999a)]{vanderMarel}	van der Marel, R. P. 1999, AJ 117, 744
\bibitem[van der Marel(1999b)]{vanderMarel99}	van der Marel, R. P. 1999 in ASP Conference Series 182, Galactic Dynamics, ed. D. R. Merritt, M. Valluri, and J. A. Sellwood, (San Francisco: ASP)
\normalsize
\end{thebibliography}
\end{document}